# Quantum Optics Nature of the Elementary Excitations in Few-Layer WSe$_2$ Semiconductors

Ji Zhou[¶], Yitong Wang[¶], Debao Zhang, Xuguang Cao, Wanggui Ye, Xinye Fan, Jiqiang Ning[*], and Shijie Xu[*]

Department of Optical Science and Engineering, School of Information Science and Technology, Fudan University, 2005 Songhu Road, Shanghai 200438, China

[¶]These authors contributed equally to the work.

[*]Corresponding authors, emails: jqning@fudan.edu.cn; xusj@fudan.edu.cn

**Abstract**: A fully quantized description of a two-level system resonantly coupled with an electromagnetic field (light) is among the central topics of quantum electrodynamics, which is theorized by the quantum Rabi model. It is also a fundamental issue of light-matter interactions. The rapid development of two-dimensional (2D) transition-metal dichalcogenide (TMDC) atomically-thin semiconductors brings excellent great chance to test and demonstrate some basic predictions by quantum optics theory in textbook, i.e., by the quantum Rabi model. Such test and demonstration are of both scientific and technological significance because of the quick emergence of the second quantum revolution. In this Letter, we show a quantum optics demonstration of the variable-temperature optical responses of the elementary excitations in few-layer WSe$_2$ flakes. It is unraveled that the variable-temperature reflectance and fluorescence patterns of the elementary excitations (i.e., the band-edge excitons) of monolayer, bilayer and hBN capped WSe$_2$ match well with the predictions by the quantum Rabi model under the rotating wave approximation. Decoherence times, Rabi frequencies, and transition matrix elements of the elementary excitations in these few-layer WSe$_2$ flakes are found to be all negatively correlated with temperature, and to show dependence on layer number and capping layer. These findings may provide a novel perspective for comprehending the fundamental quantum physical properties of two-dimensional materials.

With the rapid advancement of two-dimensional (2D) materials, transition metal dichalcogenides like tungsten diselenide (WSe$_2$), have emerged as the forefront of scientific investigation due to their unique physical, chemical and quantum-mechanical properties.[1-3] Few-layer $WSe_2$, characterized by its layered structure and bandgap features, demonstrates significant potential for applications in the fields of optoelectronics[4], photonics[5] and emerging quantum technology[6, 7]. Unlike conventional semiconductor materials, few-layer $WSe_2$ exhibits a significant reduction in screening effects due to its spatial limitation, which leads to a more pronounced Coulomb interaction between electrons and holes, thereby generating strong exciton effects.[8] The effects are particularly evident in the reflectance and photoluminescence (PL) spectra of excitons. In most cases, the Lorentz oscillator model is used to describe the behavior of electrons in semiconductors, and satisfactory results have been achieved with the classical Lorentz model.[9] However, for 2D materials composed of only a few atomic layers, quantum effects become particularly pronounced, leading to significant differences between the traditional macroscopic Lorentz model and observed phenomena. Therefore, to accurately describe the characteristics of these materials, it is imperative to introduce the theoretical framework of quantum optics. In this work, we emphasize the quantum Rabi theory of light-matter interaction in a two-level system and successfully apply it to quantitatively explain the optical spectral characteristics of the elementary excitations (e.g., the band-edge A excitons) in the few-layer $WSe_2$ materials at various temperatures, revealing the mechanism by which decoherence time influences these phenomena.[10]

Monolayer and bilayer $WSe_2$ flakes were prepared from a high-quality bulk crystal through mechanical exfoliation method, initially identified using an optical microscope, and subsequently verified through PL and Raman spectroscopic measurements. The prepared samples were transferred onto a silicon substrate using a dry transfer method, and then sent to a home-designed and assembled multi-function integrated micro-spectroscopic system for in-situ variable-temperature PL and reflection spectral measurements.[11] This system is connected with a helium closed-cycle cryostat providing a varying temperature range from 5 K to 300 K. The reflectance and PL spectra (noisy curves) of the monolayer and bilayer $WSe_2$, measured at 5 K under the excitation of a 532 nm laser are shown in Figures 1a and 1b, respectively. In the reflectance spectrum (upper panel), a characteristic peak is observed near 1.75 eV, which is assigned to the A exciton resonance, consistent with the peak position reported in the literature.[12] The reflectance structures of both samples exhibit a typical Fano line shape, with the monolayer showing enhanced oscillation amplitude over the bilayer. The PL spectra (lower panel) consist of multiple peaks, among which the emission of the A excitons is located at the highest energy position. The PL peak position of the A excitons is well consistent with their reflectance spectral structure.[13] Remaining PL peaks are considered to be the emissions of trions and defect-bound excitons.[14, 15] The low-temperature PL spectrum of the bilayer sample shows a weak direct transition emission and a broad indirect emission (partly shown). Compared with the monolayer sample, the bilayer has a lower emission energy and weaker intensity mainly due to its indirect band gap nature, as argued below. Figure 1c depicts the band structure schematics of monolayer and bilayer $WSe_2$. Monolayer $WSe_2$ possesses a direct bandgap with both valence band maximum (VBM) and conduction band minimum (CBM) located at the K point in the Brillouin zone. In starp contrast, the bilayer $WSe_2$ possesses an indirect bandgap, with the VBM and CBM located at different momentum points, leading to a significant reduction in PL intensity.[16, 17]

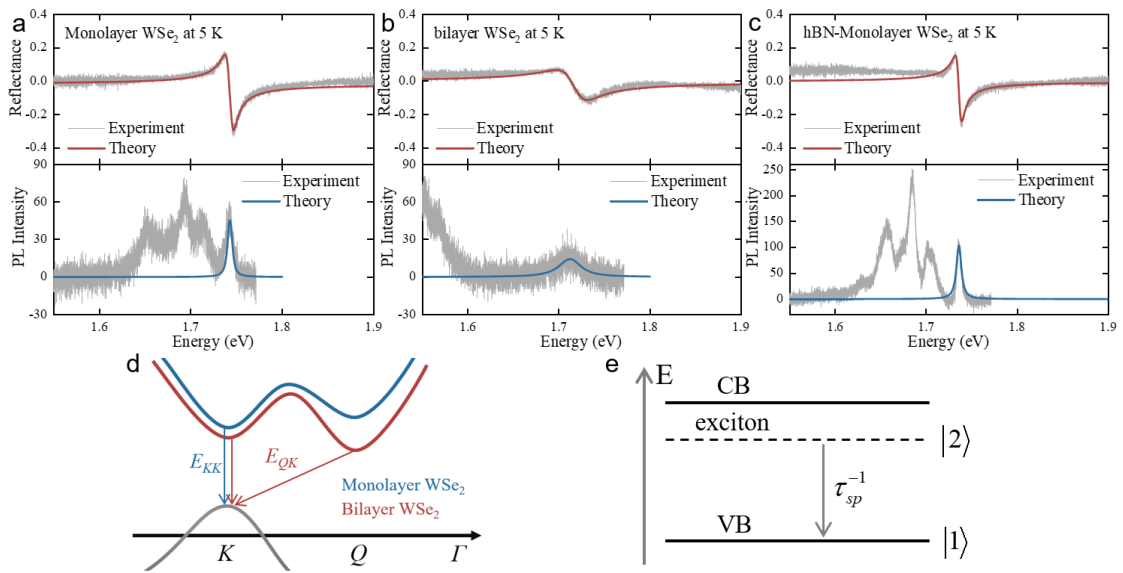

**Figure 1.** Experimental (noisy curves) and theoretical (smooth lines) reflectance spectra (upper panel) and PL spectra (lower panel) of monolayer **a**, bilayer **b**, and hBN-monolayer $WSe_2$ **c** at 5 K.

**d** Schematic diagram of the conduction and valence bands of the monolayer and bilayer $WSe_2$ with A exciton transitions. **e** Diagram of a two-level quantum system for excitons.

As seen in Figure 1, the reflectance spectra of the monolayer, bilayer, and hBN-capped $WSe_2$ exhibit intriguing excitonic resonance structures. Here, we attempt to compute them by employing the two-level quantum Rabi model of light-matter interaction described in MIT's textbook.[10] As schematically shown in Figure 1d, the band-edge A exciton can be regarded as a two-level quantum system. In such system, the Bloch equations describing its interaction with an external electromagnetic field can be represented as follows:[10]

$$\begin{cases} \dot{d} = \left(j\omega_{eg} - \frac{1}{2\tau_{sp}}\right)d - j\Omega_r^* e^{j\omega t} w \\ \dot{w} = -\frac{w+1}{\tau_{sp}} - 2j\Omega_r e^{-j\omega t} d + 2j\Omega_r^* e^{j\omega t} d^* \end{cases}, \quad (1)$$

where $d$ represents the dipole moment of a single dipole oscillator, $w$ is the population difference between the upper and lower energy levels, $\tau_{sp}$ is the lifetime of the upper energy level, $\omega_{eg}$ is the transition frequency and $\omega$ is the frequency of the external electric field. $\Omega_r = \vec{M} \cdot \vec{e}_p^* E_0^* / 2\hbar$ is the Rabi frequency, where $\vec{M}$ is the transition matrix element, $\vec{e}_p$ is the polarization unit vector of the external electric field, and $E_0$ is the amplitude of the external electric field. Under the coherent action of a resonant electromagnetic field, a two-level quantum system undergoes sinusoidal oscillation between the two energy levels at this frequency.[18] Assuming the dipole moment is in resonance with the external electric field, i.e., $d = d_0 e^{j\omega t}$, and $w$ is independent of time with a value of $w_s$, the equation set (1) can be solved as follows:

$$d_0 = -\frac{j}{2\hbar} \frac{(\vec{M}^* \vec{e}_p) w_s}{\frac{1}{2\tau_{sp}} + j(\omega - \omega_{eg})} E_0 \; ; \quad w_s = -\frac{1}{1 + \frac{I}{I_s} L(\omega)}, \quad (2)$$

where $L(\omega) = \frac{\left(\frac{1}{2\tau_{sp}}\right)^2}{\left(\frac{1}{2\tau_{sp}}\right)^2 + (\omega - \omega_{eg})^2}$ is the Lorentzian function related to the lifetime, $I_s = \left(\frac{4\tau_{sp}^2 Z_F}{\hbar^2} |\vec{M}^* \cdot \vec{e}_p|^2\right)^{-1}$ is the saturation intensity, and $I = \frac{1}{2Z_F}|E_0|^2$ is incident light intensity, here $Z_F$ is the wave impedance of the medium. It should be noted that there exists a calculation or type error in the MIT textbook, resulting in a missing negative sign in front of $d_0$.

The expectation value of the dipole moment is $\langle d \rangle = -(\vec{M}^* d + \vec{M} d^*)$, and the complex polarization of the medium is $\vec{P}_0 = -2N\vec{M}d_0^*$, where $N$ is the number density of atoms per unit volume. It should be noted that the expectation value of the dipole moment is a real number, while the complex polarization involves phase relationships. Considering that the dipole moment behaves as a forced oscillation under the action of an external electric field, its phase should lag that of the external electric field. Therefore, we incorporate expectation value of the latter half of the dipole moment into the calculation of the complex polarization rate, which differs from the original treatment. The final expression for the complex dielectric permittivity is hence obtained:

$$\tilde{\varepsilon}(\omega) = \varepsilon_b - |\vec{M}| \frac{jN}{\hbar \varepsilon_0} \frac{w_s}{\frac{1}{2\tau_{sp}} + j(\omega_{eg} - \omega)}, \qquad (3)$$

where $\varepsilon_b$ is the background dielectric constant.[19] Starting from this formula, the reflection and extinction coefficients of the material can be calculated using the following expressions:

$$R = \left| \frac{\sqrt{\tilde{\varepsilon}(\omega)} - 1}{\sqrt{\tilde{\varepsilon}(\omega)} + 1} \right|^2 ; \qquad \kappa = Im\left(\sqrt{\tilde{\varepsilon}(\omega)}\right). \qquad (4)$$

The influence of various parameters including the background dielectric constant $\varepsilon_b$, the lifetime $\tau_{sp}$, the transition matrix element $|\vec{M}|$, and the external field intensity $I$, on the reflectivity in the model was obtained, as depicted in Figure 2. Under the approximation of $\varepsilon_b = 1$ (no background polarization), the theoretical reflectivity exhibits a symmetric Lorentzian line shape (not shown), which is a typical case; as $\varepsilon_b$ increases, the reflectivity gradually evolves into a quasi-Fano line shape, with the reflectivity in the non-resonant region rising synchronously. Figure 2b illustrates the theoretical reflectivity for different lifetimes of the upper level. With increasing lifetime, the reflectivity oscillations exhibit enhanced amplitude, while the resonance structures on the reflectivity curve become sharper and more distinct. It is evident that $\tau_{sp}$ is the primary factor affecting the slope of the intermediate spectral region in the reflection structure, thereby enabling lifetime quantification via gradient analysis. The lifetime in this context corresponds to the decoherence time of the two-level quantum system, characterizing the time scale over which the coherence of the quantum state gradually diminishes due to interactions with the external environment.[20] Figure 2c shows the modulation of reflectivity by the transition matrix element, which proportionally enhances the oscillation amplitude of the reflection structure. Figure 2d illustrates the dependence of theoretical reflectivity on incident optical field intensity. The amplitude of the reflected oscillation exhibits an inverse dependence on the external field intensity. This behavior is distinct from the patterns observed with the other parameters: It neither keeps symmetry of the resonance structure around the center (as lifetime does) nor scales proportionally (as matrix element does). Specifically, the enhanced external optical field intensity exhibits asymmetric modulation effects, with significantly stronger suppression of the left peak compared to the relatively minor influence on the right valley.

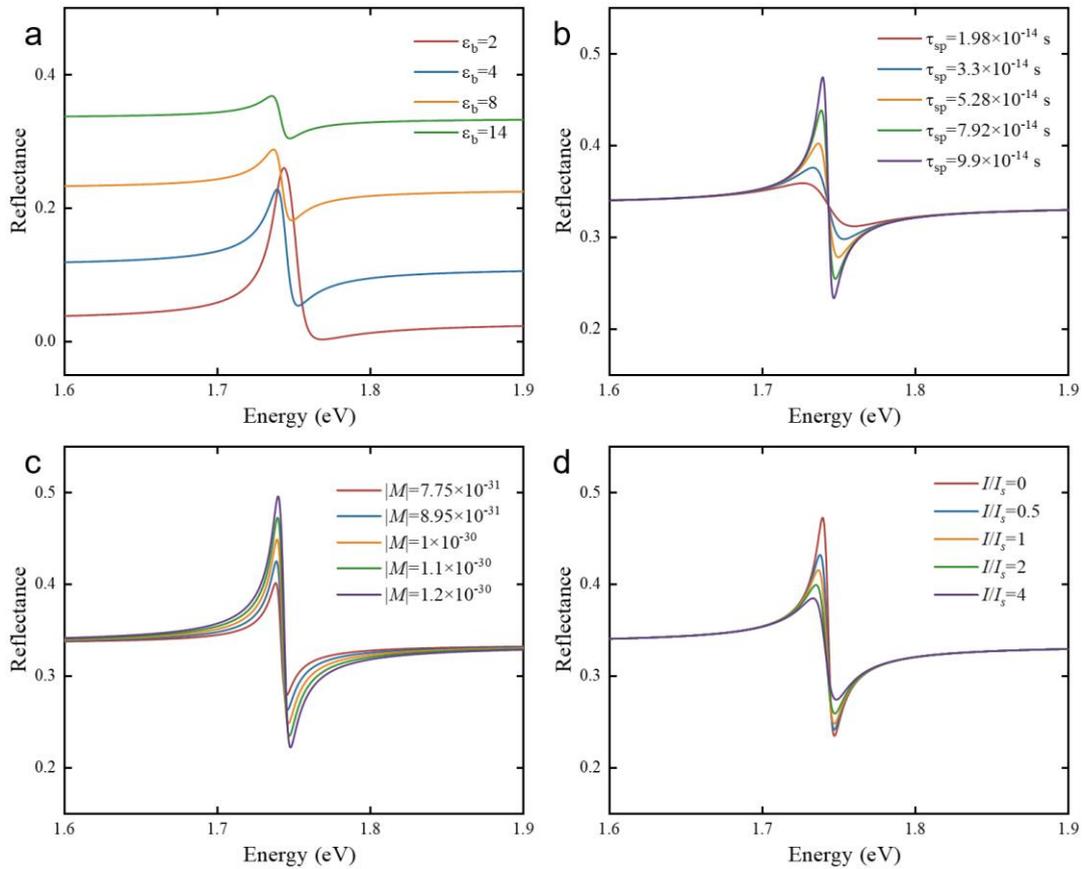

**Figure 2.** Parameter-dependent modulation of reflection spectrum. **a** Background dielectric constant.

**b** Lifetime of the upper level. **c** Transition matrix element. **d** External optical field intensity.

Now let us return to inspect the experimental reflectance spectra shown in the upper panel of Figure 1a and b. The solid lines represent the theoretical reflectance spectra of the monolayer and bilayer $WSe_2$, calculated with the parameters' values listed in Table 1. According to the von Roosbroeck-Shockley relation, in an excitonic quasi-two-level system, for narrow wavelength spans, the extinction coefficient and PL intensity exhibit identical line shape, distinguished solely by their peak positions and relative amplitudes.[21] Consequently, we utilize the extinction coefficient to model the PL spectra, employing a slightly reduced excitation power ($I$) compared to reflectance spectral calculation, while other parameters remain unchanged. The theoretical curves (smooth lines) are shown in the lower panel of Figure 1a and b. In both reflectance and PL spectra, the theoretical curves show good consistency with the experimental data. The bilayer sample exhibits a lower transition frequency $\omega_{eg}$, indicating the intrinsic bandgap reduction in bilayer systems (Figure 1c). It is worth noting that the bilayer sample has a much shorter decoherence time, which may be attributed to the interlayer coupling accelerating the loss of quantum coherence.[22] In bilayer transition metal dichalcogenides, interlayer coupling introduces additional scattering channels that can disrupt the coherence of excitons. For example, interlayer coupling facilitates interband charge transfer through tunneling between adjacent layers, and these transition processes introduce random phase changes, thereby shortening the decoherence time. In addition, interlayer coupling also promotes energy transfer of excitons between adjacent layers, which further accelerates the decoherence process. Conversely, larger transition matrix elements are observed in the bilayer sample. According to the Fermi's Golden Rule, the magnitude of the transition matrix element is directly related to the rate of the transition process and is proportional to the overlap between the wave functions of the initial and final states.[23] In bilayer structures, interlayer coupling not only alters the energy level structures of electron wavefunctions and exciton states, but gives rise to interlayer excitons which are absent in monolayers. These distinctions reduce the wavefunction overlap integral, which in turn influences the transition rates.

**Table 1.** Values of the parameters used in the calculations of the reflection spectra for the monolayer, bilayer, and hBN-capped WSe$_2$ samples at 5 K.

| | $\omega_{eg}$ (eV) | $\tau_{sp}$ ($\times 10^{-13}$ s) | $M$ ($\times 10^{-30}$) | $\varepsilon_b$ |
|---|---|---|---|---|
| Monolayer WSe$_2$ | 1.7412 | 2.125 | 1.265 | 14.5 |
| Bilayer WSe$_2$ | 1.7119 | 0.601 | 2.272 | 14.5 |
| hBN-capped WSe$_2$ | 1.7350 | 2.719 | 1.039 | 14.5 |

To investigate the quantum optical responses of the materials at different temperatures, the variable-temperature reflectance and PL spectra of the monolayer and bilayer WSe$_2$ were measured in a wide range of 5-300 K, as shown in Figure S1. From the reflectance spectra, it can be observed that with increasing the temperature, the resonance structures of the A and B excitons in both samples exhibit redshift with varying degrees, accompanied by a decay in amplitude. Due to the effects of dimensionality reduction and screening, under the strict 2D confinement conditions, the binding energy of excitons is significantly enhanced, allowing exciton-related phenomena to be observed even at room temperature.[24] Figure S1c shows the PL spectra of the monolayer sample at different temperatures. At low temperatures, the emissions of both free excitons and localized excitons coexist, but they are highly responsive to temperature. As the temperature increases, the localized excitons become thermalized and transform into free excitons, leading to a decrease in the intensity of the localized exciton emission. As a result, the emissions of the localized excitons become undetectable at temperatures above 220 K.[12] Meanwhile, the emission intensity of the free excitons is positively correlated with temperature for temperatures below 90 K. Of course, when the temperature is beyond 90 K, the intensity of the free excitons turns to decline due to the thermal quenching effect, accompanied with a noticeable broadening. In contrast, the bilayer sample shows weaker localized exciton emissions at low temperatures, instead of an indirect bandgap broad emission near 1.55 eV. The A exciton emission in the bilayer is much weaker and broader. As the temperature rises, the A exciton emission exhibits a noticeable redshift, while the indirect emission undergoes an initial blueshift followed by a redshift.

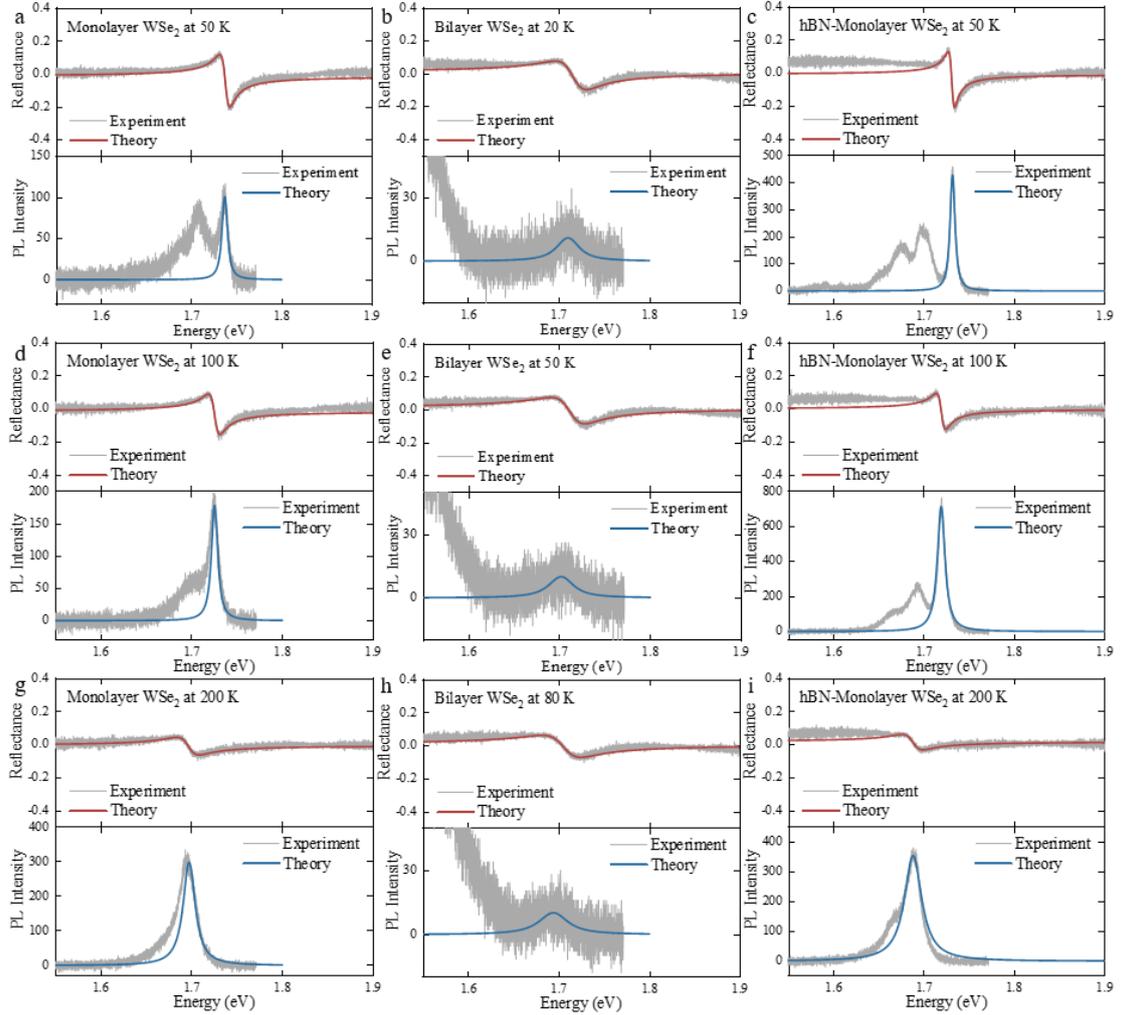

**Figure 3.** Experimental (noisy curves) and theoretical (smooth lines) reflectance spectra (upper panel) and PL spectra (lower panel) of the monolayer, bilayer, and hBN-capped $WSe_2$ at various temperatures.

Figure 3 presents several representative variable-temperature reflectance and PL spectra (noisy curves) and the corresponding theoretical spectra (smooth solid lines) for the monolayer, bilayer, and hBN-capped $WSe_2$ samples. Decoherence times and transition matrix elements determined in the calculations are plotted as a function of temperature in Figures 4a and 4c. The decoherence times of both samples are negatively correlated with temperature. Specifically, the decoherence time of the monolayer material is much longer and more significantly dependent on temperature compared with the bilayer sample. In addition to the radiative recombination processes, the decoherence time of two-dimensional materials can be affected by other physical processes including non-radiative recombination and phonon scattering, which may be given by $\frac{1}{\tau_{sp}} = \gamma_{rad} + \gamma_{K-K} + \gamma_{K-K'} + \gamma_{K-A}$, where $\gamma_{rad}$ is the temperature-independent radiative decay rate, $\gamma_{K-K}$ accounts for the intravalley exciton-phonon scattering, and $\gamma_{K-K'}$ and $\gamma_{K-A}$ are for the inter-valley scatterings.[25] At low temperatures, the radiative recombination is the dominant decoherence mechanism. As the temperature increases, the phonon scatterings gradually become the major decoherence mechanism. As a result, the decoherence of both monolayer and bilayer samples become quite efficient at high temperatures. In most cases, the matrix element is considered as a constant. However, here the matrix element exhibits a negative correlation with temperature, which is more pronounced in the monolayer sample. As the temperature increases, the substrate exerts tensile stress on the 2D material due to the thermal expansion, causing some increase in the atomic spacing, thereby reducing the degree of overlap of the wavefunctions. Figure 4b presents the Rabi frequencies obtained in the calculations versus temperature. According to the definition of the Rabi frequency, when the external field strength remains constant, the Rabi frequency is directly proportional to the magnitude of the transition matrix element. The calculation shows that the Rabi frequency is five orders of magnitude lower than the excitation light frequency, which is consistent with the assumptions made in the derivation of Equation (2), namely that the population of particles can be considered to be time-independent. It is worth noting that the Rabi frequency of the bilayer sample is obviously larger than that of the monolayer sample. Moreover, the Rabi frequency of the bilayer is dependent weakly on temperature, while the Rabi frequency of the monolayer shows a noticeable dependence on temperature.

Figure S2 illustrates the population difference $w_s$ between the upper and lower energy levels for different values of $I/I_s$. According to the quantum Rabi theory, when a two-level system resonates with an external field, the population of particles oscillates at the Rabi frequency. If there is a detuning, i.e., when $\Delta\omega$ is large enough, there have no interactions between the system and the external field, and all particles remain at the ground state. That means $w_s = -1$. However, when the external optical field resonates with the two-level system, i.e., when $\omega = \omega_{eg}$, the Rabi oscillation intensity is positively related with the external field intensity. There exists a periodic energy exchange between the light field and the system, and the average value of the population difference between the two levels tends to approach zero.

Monolayer WSe$_2$ samples with a hBN capping (Sample 1) and a hBN basing layer (Sample 2) were also analyzed, respectively, as shown in Figure S3 in supporting information. Overall, the properties of both samples are closer to those of the pristine monolayer WSe$_2$. Specifically, their bandgaps are nearly identical and slightly smaller than that of the pristine monolayer WSe$_2$ (see Figure S3a in supporting information). Figure S3b shows the differences in decoherence times among the four samples. The values and temperature-dependent trend of Sample 2 are almost identical to those of the pristine monolayer WSe$_2$, indicating minimal influence of the substrate material on the decoherence time of monolayer WSe$_2$. In contrast, below 70 K, the decoherence time of Sample 1 exhibits negligible temperature dependence, significantly differing from the behavior of the pristine monolayer WSe$_2$; above 100 K, however, its decoherence time converges with that of the pristine monolayer WSe$_2$. Figure S3c compares the obtained transition matrix elements of the samples. Evidently, they follow a decreasing sequence: hBN-based > pristine WSe$_2$ > hBN-capped. Regarding temperature dependence, the transition matrix elements of the hBN-based and hBN-capped WSe$_2$ exhibit a fluctuating decrease and stabilize at higher temperatures (>180 K).

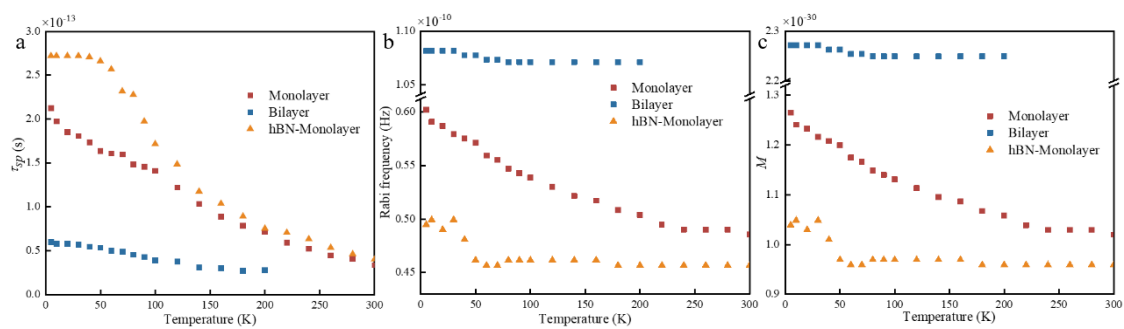

**Figure 4.** Comparison of the determined decoherence times (**a**), Rabi frequencies (**b**), and matrix

elements (**c**) for the monolayer (brown squares), bilayer (blue squares), and hBN-capped monolayer WSe$_2$ (upper triangles).

In summary, the variable-temperature reflectance and PL spectra of the monolayer, bilayer, and hBN-capped WSe$_2$ samples were measured and theoretically explained with the two-level quantum Rabi model for light-matter interactions in textbooks. Additionally, the impact of various key parameters on reflectivity of few-layer 2D semiconductors were calculated and analyzed. The decoherence time and matrix elements of the elementary excitations in few-layer WSe$_2$ were determined as a function of temperature. At cryogenic temperatures, the decoherence time of the bilayer WSe$_2$ was found to be ~60 fs, which is much shorter than ~210 fs of the monolayer sample, while a hBN capping layer can significantly extend the decoherence time of the elementary excitations in monolayer WSe$_2$. However, the matrix element and Rabi frequency of the bilayer sample were evidently larger, while those parameters of the hBN-capped sample were smallest. These key quantities all show negative dependence on temperature, but dependence depends largely on layer number and capping layer. Those parameters of the monolayer sample exhibit a relatively much stronger dependence on temperature. The experimental demonstration of quantum optics nature of the elementary excitations in ultimate 2D semiconductors represents an important progress in solid quantum optics, and open a way for investigating intriguing properties of various elementary excitations including dark excitons in various 2D semiconductors.[26]

Acknowledgments: The work was financially supported by the National Natural Science Foundation of China (No. 12074324).

Supporting Information

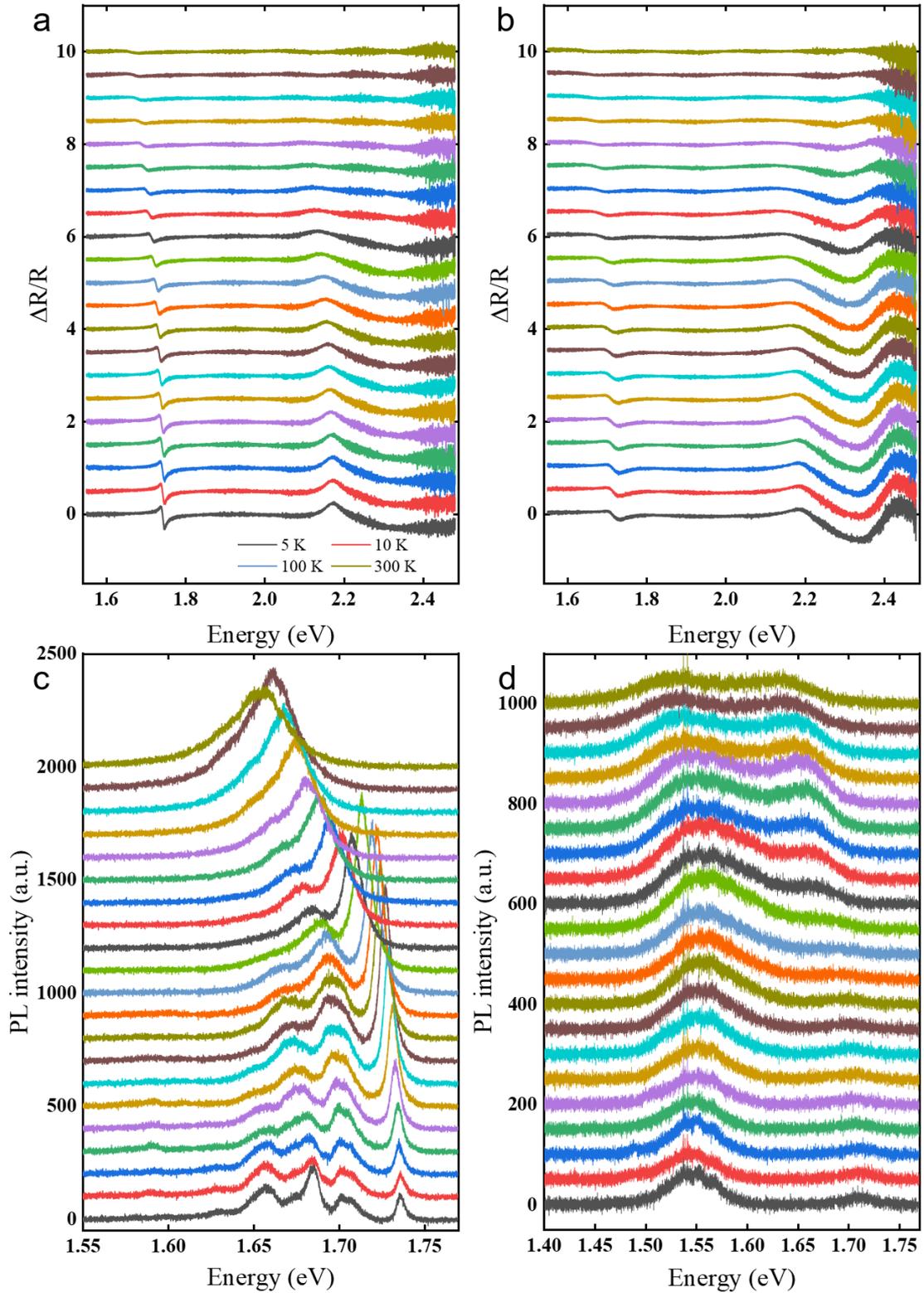

**Figure S1. a, c** Temperature-dependent reflectance and PL spectra of the monolayer WSe$_2$ sample. **b, d** Temperature-dependent reflectance and PL spectra of the bilayer WSe$_2$ sample. The data in the figure are plotted at interval of 10 K in the temperature

range of 10-100 K, and at interval of 20 K above 100 K.

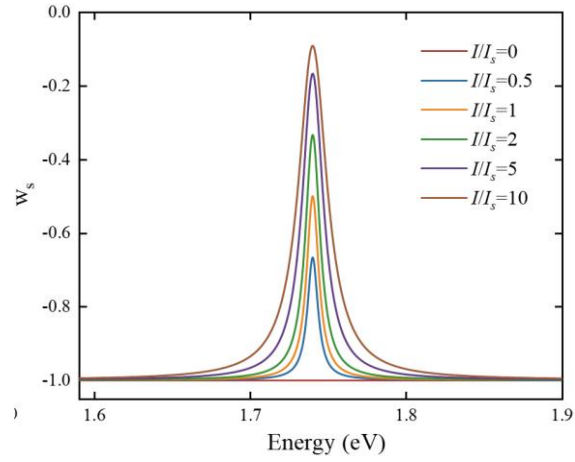

**Figure S2.** Population difference between the upper and lower energy levels vs. relative intensity of the external field.

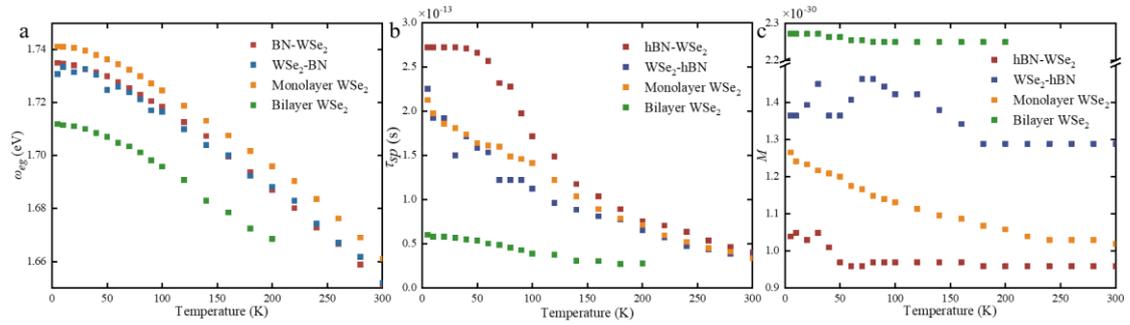

**Figure S3.** Comparison of four WSe$_2$ configurations: i) monolayer WSe$_2$ with hBN capping layer; ii) monolayer WSe$_2$ with hBN underlying layer; iii) monolayer WSe$_2$; iv) bilayer WSe$_2$. The figure shows: **a** band gaps; **b** decoherence times; **c** transition matrix elements.

1. Zhao, W., et al., *Evolution of electronic structure in atomically thin sheets of $WS_2$ and $WSe_2$.* ACS Nano, 2013. **7**(1): p. 791-797.
2. Hao-Zhe, Z., et al., *Research progress of two-dimensional transition metal dichalcogenide phase transition methods.* Wuli, 2020. **69**: p. 246101.
3. Choi, W., et al., *Recent development of two-dimensional transition metal dichalcogenides and their applications.* Materials Today, 2017. **20**(3): p. 116-130.
4. Cao, G., et al., *Optoelectronic investigation of monolayer $MoS_2$/$WSe_2$ vertical heterojunction photoconversion devices.* Nano Energy, 2016. **30**: p. 260-266.
5. Guo, Q., et al., *Boosting exciton transport in $WSe_2$ by engineering its photonic substrate.* ACS Photonics, 2022. **9**(8): p. 2817-2824.
6. Yi, Y., et al., *Recent advances in quantum effects of 2D materials.* Advanced Quantum Technologies, 2019. **2**(5-6): p. 1800111.
7. Chen, P., et al., *Large quantum-spin-Hall gap in single-layer 1T' $WSe_2$.* Nature Communications, 2018. **9**(1): p. 1-7.
8. Massicotte, M., et al., *Dissociation of two-dimensional excitons in monolayer $WSe_2$.* Nature Communications, 2018. **9**(1): p. 1633.
9. Koperski, M., et al., *Optical properties of atomically thin transition metal dichalcogenides: observations and puzzles.* Nanophotonics, 2017. **6**(6): p. 1289-1308.
10. Kärtner, F., *Fundamentals of Photonics: Quantum Electronics*. 2006, Massachusetts Institute of Technology: MIT, OpenCourseWare.
11. Cao, X., et al., *Dielectric screening effects in the decoherence of excitons and exciton-phonon scattering in atomical monolayer $WS_2$ semiconductors.* Physical Review B, 2025. **112**(4): p. 045420.
12. Yan, T., et al., *Photoluminescence properties and exciton dynamics in monolayer $WSe_2$.* Applied Physics Letters, 2014. **105**(10): p. 101901.
13. Hu, H., et al., *Probing angle-resolved reflection signatures of intralayer and interlayer excitons in monolayer and bilayer $MoS_2$.* Nano Research, 2023. **16**(5): p. 7844-7850.
14. Wu, Z., et al., *Defect activated photoluminescence in $WSe_2$ monolayer.* The Journal of Physical Chemistry C, 2017. **121**(22): p. 12294-12299.
15. Jadczak, J., et al., *Probing of free and localized excitons and trions in atomically thin $WSe_2$, $WS_2$, $MoSe_2$ and $MoS_2$ in photoluminescence and reflectivity experiments.* Nanotechnology, 2017. **28**(39): p. 395702.
16. Glazov, M., et al., *Excitons in two-dimensional materials and heterostructures: Optical and magneto-optical properties.* MRS Bulletin, 2024. **49**(9): p. 899-913.
17. Zeng, H. and X. Cui, *An optical spectroscopic study on two-dimensional group-VI transition metal dichalcogenides.* Chemical Society Reviews, 2015. **44**(9): p. 2629-2642.
18. Rabi, I.I., et al., *The molecular beam resonance method for measuring nuclear magnetic moments.* Physical Review, 1939. **55**(6): p. 526.
19. Fox, M., *Optical Properties of Solids*. Vol. 3. 2010: Oxford University Press.
20. Arzano, M., V. D'Esposito, and G. Gubitosi, *Fundamental decoherence from

   *quantum spacetime.* Communications Physics, 2023. **6**(1): p. 242.
21. Van Roosbroeck, W. and W. Shockley, *Photon-radiative recombination of electrons and holes in germanium.* Physical Review, 1954. **94**(6): p. 1558.
22. Helmrich, S., et al., *Phonon-assisted intervalley scattering determines ultrafast exciton dynamics in MoSe$_2$ bilayers.* Physical Review Letters, 2021. **127**(15): p. 157403.
23. Fermi, E., *Quantum theory of radiation.* Reviews of Modern Physics, 1932. **4**(1): p. 87.
24. Lin, Y., et al., *Dielectric screening of excitons and trions in single-layer MoS$_2$.* Nano Letters, 2014. **14**(10): p. 5569-5576.
25. Selig, M., et al., *Excitonic linewidth and coherence lifetime in monolayer transition metal dichalcogenides.* Nature Communications, 2016. **7**(1): p. 13279.
26. Cao, X., et al., *Revealing decoherence dependence of the bright A excitons on the energy positions of the intervalley dark excitons in WS$_2$ semiconductors.* To be published in ACS Nano.